\documentclass[letterpaper]{article} 
\usepackage[]{aaai25}  
\usepackage{times}  
\usepackage{helvet}  
\usepackage{courier}  
\usepackage[hyphens]{url}  
\usepackage{graphicx} 
\usepackage{natbib}  
\usepackage{caption} 
\usepackage{algorithm}
\usepackage{algorithmic}

\usepackage{newfloat}
\usepackage{listings}
\DeclareCaptionStyle{ruled}{labelfont=normalfont,labelsep=colon,strut=off} 
\floatstyle{ruled}
\newfloat{listing}{tb}{lst}{}
\floatname{listing}{Listing}
\nocopyright 

\title{Transient Multi-Agent Path Finding for\\Lifelong Navigation in Dense Environments}
\author {
    Jonathan Morag,
    Noy Gabay,
    Daniel Koyfman,
    Roni Stern
}
\affiliations {
    Ben-Gurion University of the Negev \\
    \{moragj, noygab, koyfdan\}@post.bgu.ac.il, roni.stern@gmail.com
}

\usepackage{xspace}
\usepackage{mathtools}
\usepackage{amsthm}
\usepackage{amsfonts}
\usepackage{booktabs}

\newcommand{\soc}{SOC\xspace}
\newcommand{\len}{len\xspace}
\newcommand{\mkspn}{Makespan\xspace}
\newcommand{\prp}{PrP\xspace}
\newcommand{\prpt}{PrPt\xspace}
\newcommand{\lnst}{LNSt\xspace}

\newcommand{\lacam}{LaCAM\xspace}
\newcommand{\lacamt}{LaCAMt\xspace}

\newcommand{\classic}{Classical\xspace}
\newcommand{\cmapf}{MAPF\xspace}
\newcommand{\astar}{A$^*$\xspace}
\newcommand{\astart}{A$^*_t$\xspace}
\newcommand{\tmapf}{TMAPF\xspace}

\newcommand{\maze}{maze-128-128-2\xspace}
\newcommand{\warehouse}{warehouse-20-40-10-2-1\xspace}
\newcommand{\emptymap}{empty-48-48\xspace}

\newcommand{\commentout}[1]{}
  
\newcommand{\anomaly}{MAPF-LMAPF mismatch\xspace}
\newcommand{\planningFrequency}{5\xspace}
\newcommand{\responseTime}{5\xspace}

\begin{document}

\maketitle

\begin{abstract}
Multi-Agent Path Finding (MAPF) deals with finding conflict-free paths for a set of agents from an initial configuration to a given target configuration. 
The Lifelong MAPF (LMAPF) problem is a well-studied online version of MAPF in which an agent receives a new target when it reaches its current target. 
The common approach for solving LMAPF is to treat it as a sequence of MAPF problems, periodically replanning from the agents' current configurations to their current targets. 
A significant drawback in this approach is that in MAPF the agents must reach a configuration in which all agents are at their targets simultaneously, which is needlessly restrictive for LMAPF. 
Techniques have been proposed to indirectly mitigate this drawback. We describe cases where these mitigation techniques fail. 
As an alternative, we propose to solve LMAPF problems by solving a sequence of modified MAPF problems, in which the objective is for each agent to eventually visit its target, but not necessarily for all agents to do so simultaneously. We refer to this MAPF variant as Transient MAPF (TMAPF) and propose several algorithms for solving it based on existing MAPF algorithms. 
A limited experimental evaluation identifies some cases where using a TMAPF algorithm instead of a MAPF algorithm with an LMAPF framework can improve the system throughput significantly. 

\end{abstract}

\section{Introduction}

The Multi-Agent Path Finding (MAPF) problem deals with finding conflict-free paths for a set of agents from an initial configuration to a given target configuration. 
MAPF problems manifest in real-world applications such as automated warehouses~\cite{Ma2017,li2021lifelong,morag2023adapting}, and a wide range of MAPF algorithms have been proposed~\cite{felner2017search}. 
In many MAPF applications, it is often necessary to solve an online version of MAPF called 
\emph{Lifelong MAPF (LMAPF)}~\cite{li2021lifelong,vsvancara2019online}, where when an agent reaches its target, it is assigned a new target. 
LMAPF has attracted significant interest in both academia and industry, 
including a recent LMAPF competition that was sponsored by Amazon Robotics.\footnote{https://www.leagueofrobotrunners.org} 
The common approach for solving LMAPF problem is to treat it as a sequence of classical MAPF problems, where each MAPF problem deals with finding paths from the agents' current configuration to their up-to-date targets. 

\begin{figure}[t]
      \centering
      \includegraphics[width=0.4\columnwidth]{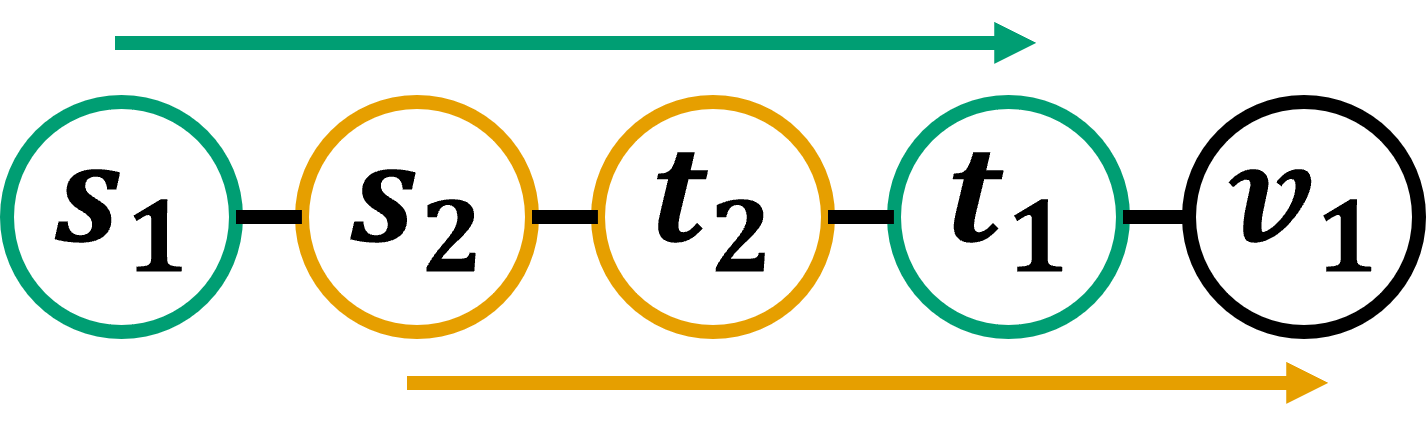}
      \caption{LMAPF problem that cannot be solved as a sequence of MAPF problems.}
       \vspace{-0.4cm}
      \label{fig:SolvableUnsolvableExample}
\end{figure}

A significant drawback in this approach stems from the fact that, unlike in MAPF, the agents in LMAPF are not required to reach a configuration where all agents are at their targets, since once an agent reaches its target it may receive a new one. 
For example, consider the LMAPF problem in Figure~\ref{fig:SolvableUnsolvableExample}. 
Two agents $a_1$ and $a_2$ are positioned in vertices $s_1$ and $s_2$ and assigned target $t_1$ and $t_2$, respectively.  
As an LMAPF problem, both targets can be reached: once $a_2$ reaches its target $t_2$ it receives a new target and can choose to move to $v_1$ so that $a_1$ can also reach its target. 
Solving this LMAPF problem as a sequence of MAPF problems, however, will fail, because as a MAPF problem, this problem has no solution as the agents cannot swap positions in this graph.

\citet{varambally2022mapf} and others have acknowledged this MAPF-LMAPF ``mismatch'' and proposed several techniques to mitigate it to some extent. These techniques include planning for a limited horizon~\cite{li2021lifelong}, dynamic replanning whenever a target is reached~\cite{varambally2022mapf}, 
defining ``dummy paths'' to follow after visiting a target ~\cite{liu2019task}, 
and assigning each agent with a sequence of targets~\cite{grenouilleau2019multi}. 
We discuss these techniques and show they do not provide a complete solution to the \anomaly .

As an alternative, we propose to solve LMAPF problems by solving a sequence of modified MAPF problems that we call \emph{Transient MAPF} (TMAPF). 
In TMAPF, the objective is to find conflict-free paths for the agents such that each agent visits its target at some point along its path, but there is no need to reach a configuration in which all agents are at their targets at the same time. 
We discuss the properties of the TMAPF problem and adapt several popular MAPF algorithms to solve it. 
Finally, we conduct a small-scale set of experiments comparing the throughput of an LMAPF system when using  MAPF and TMAPF algorithms. In some cases, the advantage of using TMAPF algorithms is considerable.

\section{Background}

A {\em Multi-Agent Path Finding (MAPF)} problem is defined by a graph $G=(V,E)$; a set of agents $A$; and a source and target vertex for each agent $a_i\in A$, denoted $s_i$ and $t_i$, respectively. 
A \emph{solution} $\pi$ to a MAPF problem is a mapping of each agent $a_i\in A$ to a path $\pi_i$ in $G$ such that $\pi_i$ starts at $s_i$, ends at $t_i$, and does not \emph{conflict} with any path that is mapped in $\pi$ to a different agent.
\footnote{We assume all vertices in $G$ have self-edges, allowing paths in which an agent waits at a vertex.}
In this work, we consider as a conflict only {\em vertex conflicts} and {\em swapping conflicts}~\cite{stern2019multi}. 
A pair of paths $\pi_i$ and $\pi_j$ has a vertex conflict if there exists an index $x$ such that $\pi_i[x]=\pi_j[x]$, where $\pi_i[x]$ is the $x^{th}$ vertex in the path $\pi_i$. 
A swapping conflict occurs if there exists an index $x$ such that $\pi_i[x]=\pi_j[x+1]$ 
and $\pi_i[x+1]=\pi_j[x]$.  
The length of a path is defined as the number of vertices (non-unique) in the path, minus one ($\len(\pi_i) - 1$).
This version of MAPF is known as {\em \classic MAPF}. 
\emph{Sum Of Costs} (\soc) and \emph{Makespan} are common MAPF solution cost function. 
\soc is the sum of the lengths of all paths in the solution ($\mathit{\soc}(\pi) \coloneqq \sum (\len(\pi_i) - 1) | \pi_i \in \pi)$) and makespan is the length of the longest path in the solution ($\mathit{\mkspn}(\pi) \coloneqq \max(\len(\pi_i) - 1 | \pi_i \in \pi)$). 
A solution $\pi$ is \emph{optimal} for a given MAPF problem 
 w.r.t. a given cost function if its cost is smaller than or equal to the cost of any solutions to that MAPF problem. Some MAPF algorithms, such as Conflict Based Search (CBS)~\cite{sharon2015conflict} guarantee the solution they return is optimal, while others, such as Prioritized Planning (\prp) \cite{bennewitz2001optimizing} do not.

\emph{Lifelong Multi-Agent Path Finding (LMAPF)}~\cite{li2021lifelong,morag2023adapting}
is a generalization of MAPF in which whenever an agent reaches its target, it may be assigned a new target to reach. 
Due to its online nature, an LMAPF problem is typically solved by repeatedly interleaving planning and execution. In every \emph{planning period} a path for each agent is computed. 
Then, a prefix of the planned paths is executed, and another planning period starts.
This process continues indefinitely until interrupted by the user. 
The performance of an LMAPF algorithm is commonly evaluated using \emph{throughput}, which counts how many times agents reached their targets before a fixed number of time steps have passed. 
As noted above, a common approach for solving LMAPF is by iteratively calling a MAPF solver every planning period to find paths for the agents from their current locations to their current targets. 
Different MAPF solvers have been used for this purpose, including optimal and suboptimal MAPF algorithms.

\section{MAPF-LMAPF Mismatch and Mitigations}
The role of an LMAPF algorithm is to output plans for the agents in every planning period.
We say that an LMAPF algorithm is \emph{complete} if it does so, and incomplete otherwise. 
The example in Figure~\ref{fig:SolvableUnsolvableExample} shows that the common approach of solving LMAPF as a sequence of MAPF problems yields  incomplete LMAPF algorithms. Several techniques have been proposed for mitigating, to some extent, this incompleteness. 

\paragraph{Dynamic replanning} This means whenever a target is reached, all agents immediately replan~\cite{varambally2022mapf}. 
This option is not feasible in some cases since it requires significant computational cost. 
Moreover, this approach does not solve the problem in Fig.~\ref{fig:SolvableUnsolvableExample} as the agents will fail to find any MAPF solution. 

\paragraph{Pre-assigned sequence of goals} This means each agent is 
given a sequence of targets to reach and the LMAPF algorithm returns plans to visit the next couple of targets~\cite{grenouilleau2019multi}. 
Unfortunately, it is not always possible to know such a sequence of tasks for each agent in advance. 
Also, the planning itself is harder when planning to visit more than one node. 
Moreover, this technique is also incomplete in some cases.

\paragraph{Dummy paths} This means that after an agent reaches its target, it will follow a pre-defined path so as to avoid causing congestion near its target~\cite{liu2019task}. 
This mitigation strategy still suffers the MAPF limitation that a specific configuration of the agents must be reached. 
Indeed, identifying effective dummy paths is not trivial and poorly located dummy paths may result in an incompleteness scenario similar to Fig.~\ref{fig:SolvableUnsolvableExample}. This technique may also make planning more difficult, as it requires longer paths.

\paragraph{Limited planning horizon} 
This means ignoring conflicts that are further than $w$ time steps in the future when planning, where $w$ is a parameter. 
This technique is embodied in the Rolling Horizon Collision Resolution (RHCR) framework~\cite{li2021lifelong}, which is the state of the art in LMAPF. 
Common implementations of RHCR use LNS~\cite{li2021anytime} or \prp for planning, but any \cmapf algorithm can be easily adapted for this purpose. 
While using a limited planning horizon is common, it may lead to deadlocks and to inefficiencies stemming from myopic planning. We show such an example in Fig.~\ref{fig:snapshotBadExample}

\paragraph{Fail policies}
Morag et al.~\shortcite{morag2023adapting} proposed an LMAPF framework designed to handle planning periods in which the underlying MAPF algorithm failed to return a plan.  
They proposed several \emph{fail policies} for this purpose and techniques for utilizing partial solutions the MAPF algorithm may return. 
An LMAPF algorithm with a fail policy can be viewed as a complete LMAPF algorithm, 
yet the fail policies proposed so far are ad-hoc and may be inefficient.

In summary, while current techniques partially reduce the impact of the MAPF-LMAPF ``mismatch'', they do so indirectly and often in an incomplete manner. In this work, we propose a more direct approach to address this mismatch by modifying the type of MAPF problem being solved in every planning period. We call this modified MAPF problem \emph{Transient MAPF} (\tmapf) and define it below.

\section{A Transient Version of MAPF}

The input to a \tmapf problem is the same as the input for \cmapf . 
The solution to a \tmapf problem is also a mapping of each agent to a path in $G$. 
The key difference between \cmapf and \tmapf is that in a solution to a \cmapf problem the path $\pi_i$ mapped to each agent $a_i$ must \emph{end} at $t_i$, in a \tmapf solution it suffices that $\pi_i$ \emph{includes} $t_i$. 
One may view \tmapf as a degenerated version of LMAPF, where when an agent reaches its goal, it is not given a new goal, and use LMAPF algorithms to solve it. 
As noted above, however, most LMAPF algorithms are not designed to plan paths for agents without a goal, while this is needed for solving TMAPF problems (see Fig.~\ref{fig:SolvableUnsolvableExample}).

A notable exception to this is the Priority Inheritance with Backtracking (PIBT) algorithm~\cite{okumura2022priority}. 
PIBT was designed to solve \textit{Iterative MAPF}, which is a generalization of LMAPF in which each agent receives multiple targets to reach.  
In every iteration PIBT chooses the configuration the agents should move to in the next time step by using a sophisticated version of \prp with a one-step lookahead. 
Under certain conditions, PIBT guarantees that each agent will reach its target eventually~\cite{okumura2022priority}.

PIBT is also designed to solve \cmapf problems, by continuously re-assigning an agent to its target after it reaches it for the first time, hopefully reaching a configuration where all agents are at their targets simultaneously. 
PIBT can similarly be adapted to solve \tmapf problems by simply not re-assigning agents to their targets after reaching them, instead associating such agents with a minimal priority hereinafter, allowing other agents to move them if needed.

\subsubsection{\prp, LNS, and CBS for Transient MAPF}
The one-step lookahead planning done by PIBT often limits its effectiveness in complex scenarios with narrow corridors, which are common in automated warehouses. 
Instead, many \cmapf algorithms such as PrP, LNS, and CBS 
consider longer planning horizons, using \astar~\cite{hart1968formal} 
to find optimal or close to optimal paths 
in the search-space of locations and time~\cite{pp-old}. 
To use such algorithms for solving \tmapf problems, we propose to extend the state definition in \astar to include a Boolean flag indicating whether or not any of the state's ancestors visited the agent's target ($\langle v, x, b\rangle$). This flag is easily maintained by initializing it to $0$ when generating a child state (or the root), and changing it to $1$ if the state contains the target or if the flag was set to $1$ in the parent state ($\mathit{child.b} \leftarrow \mathit{child.v} = t_i \vee \mathit{parent.b}$).
Extending the state definition is important for two uses. First, two states that have the same time and vertex, but differ in having an ancestor that visited the target, must be considered distinct, and neither must be discarded as a duplicate of the other. Otherwise, the search may miss valid paths. Second, the definition of a goal state is modified to require that the state had an ancestor that visited the target ($\langle v, x, 1\rangle$), instead of requiring that the vertex is equal to the target. 
To differentiate between the \cmapf and \tmapf versions we call the \cmapf version \astar, and the \tmapf version \astart.

\section{Experimental Results}
\begin{figure}[t]
      \centering
      \includegraphics[width=0.8\columnwidth]{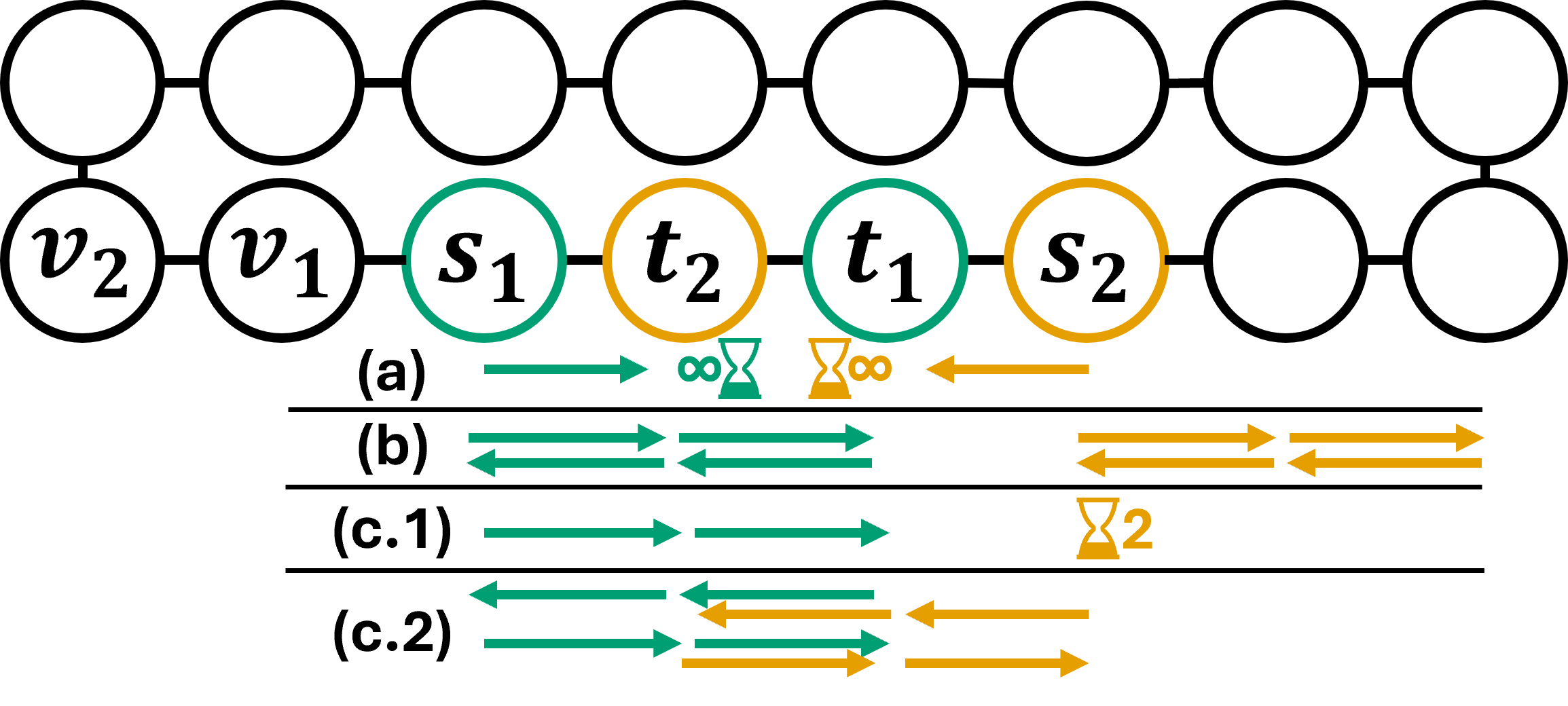}
      \caption{An LMAPF problem in which pathological behaviors occur under different conditions and solvers.}
      \label{fig:snapshotBadExample}
\end{figure}

We conducted an experimental evaluation of our \tmapf algorithms when solving LMAPF problems on standard grids from the grid-based MAPF benchmark~\cite{stern2019multi}
and on the pathological cases shown in Fig.~\ref{fig:SolvableUnsolvableExample} and~\ref{fig:snapshotBadExample}. 
In each experiment, we compare the performance of solving LMAPF problems using either a \cmapf or a \tmapf algorithm in every planning period.  
Specifically, we used PIBT, \prp , CBS, and LNS, and the \tmapf  versions of \prp , CBS, and LNS we developed, denoted as \prpt , CBSt, and \lnst . 
Our LNS implementation only used the Map-Based and Random Destroy Heuristics. 
All experiments were run on Linux virtual machines in a cluster of AMD EPYC 7763 CPUs, with 16GB of RAM each.

\paragraph{Benchmark Grids} 
This set of experiments were conducted on grids from the standard MAPF benchmark~\cite{stern2019multi}.
All algorithms were run within the RHCR~\cite{li2021lifelong} framework. 
The experiments included LMAPF problems with between 25 and 1000 agents, limited by the maximal number of agents available for each problem in the benchmark.
The number of time steps between planning iterations was set to \planningFrequency, and the planning horizon ($w$) was set to 10, a configuration which was shown to be generally effective in previous works~\cite{li2021lifelong,morag2023adapting}. The amount of (real) time for planning at each planning iteration was \responseTime seconds. The metric we measured was the throughput at time step 1000.

We generated LMAPF instances based on the grid-based MAPF benchmark, generating new targets by selecting locations in the grid randomly (uniformly). 
Consequently, multiple agents may have the same target at the same time. \tmapf algorithms handle this property organically, while \cmapf cannot handle this without planning paths where agents wait until the horizon is exceeded. 
We run all our experiments within the robust LMAPF framework described by \citet{morag2023adapting}, which includes selecting the agents that should plan in every planning period and determining a fail policy for cases where the planner fails to plan within a given time budget. 

In all maps, we observed no significant advantage for using \tmapf, i.e., \prpt exhibited similar results to \prp and LNS exhibited similar results to \lnst. 
This may be due to the limited planning horizon, which, as mentioned above, can mitigate the effect of the LMAPF-MAPF mismatch. 
To challenge our algorithms further, we repeated this LMAPF experiments but instead of generating targets randomly from any location in the map, we randomly sampled from a restricted set of 10, 20, 30, or 40 targets. 
One may argue that these experiments are more realistic than spreading the targets uniformly, especially in real-world applications of LMAPF. For instance, automated warehouses often have fixed picking stations and charging stations that would make certain locations more or less likely to appear as an agent's target. Similarly, in traffic control, locations such as public transit hubs or office buildings would be common targets for vehicles. 
Table~\ref{tbl:denseLL} shows the results of the dense target experiments on maps \emptymap and \warehouse . 
Results on other maps and numbers of agents show similar trends and are available in the supplementary materials.
For every map and number of agents, we highlighted in bold the algorithm that achieved the highest throughput between every \tmapf algorithm and its \cmapf counterpart.  
The results show that \prpt and \lnst yield a relatively higher throughput when the number of possible targets is limited. For example, the highest throughput achieved on \warehouse with 500 agents and 20 targets was 543, achieved by both \prpt and \lnst . For the same map, number of agents, and number of targets, \prp and LNS yielded a throughput of only 491 and 492, respectively. 
These results highlight the importance of using \tmapf algorithms in such hard and dense LMAPF problems.

\paragraph{Pathological Cases}
Next, we evaluated all algorithms on two special LMAPF problems. 
The first LMAPF problem we consider is based on the example depicted in Fig.~\ref{fig:SolvableUnsolvableExample}. Agent $a_1$ starts at $s_1$ and is assigned targets going back and forth between $s_1$ and $t_1$, indefinitely. The same is done for agent $a_2$ with vertices $s_2$ and $t_2$. For example, $a_1$ will be assigned targets $[t_1,s_1,t_1,s_1...]$ .
As discussed above, without limiting the planning horizon using a \cmapf solver results in planning failure, as the agents must swap their positions, which is impossible. 
Limiting the planning horizon by using RHCR helps mitigate this problem, which raises the question --- is this a substantial problem in LMAPF? 
We solved this LMAPF problem with a planning horizon of 10, running for 500 time steps, and using \prp , \prpt, and PIBT. We found that \prp and \prpt both achieved a throughput of 500, whereas PIBT achieved a throughput of 320. 
In comparing \astar expansions, \prp required a total of 13,509 expansions, whereas \prpt required only 11,965 expansions. The total runtime was 56ms for \prp , 50ms for \prpt, and 163ms for PIBT. 
This experiment demonstrates how, even when using RHCR, the new \tmapf solver, \prpt, had advantages in runtime over the \cmapf solver \prp, and in throughput over the existing \tmapf solver PIBT.

The second special LMAPF problem we consider is illustrated in 
Figure~\ref{fig:snapshotBadExample}. 
Here there are two agents, $a_1$ and $a_2$, tasked with moving repeatedly between $s_1$ and $t_1$, or $s_2$ and $t_2$, respectively.
All agents are allowed to re-plan at every time step. 
Consider a planning horizon of $1$ is used. Both the \cmapf and \tmapf solvers will behave as seen in Figure~\ref{fig:snapshotBadExample}(a), moving towards each other, and then waiting in place indefinitely, always assuming that once the planning horizon passes they will be able to swap their positions and reach their targets. This results in a throughput of 0 for both solvers. The same would happen with a horizon of $2$ for the \tmapf solver, and any horizon $\leq5$ for the \cmapf solver.
Second, assume an infinite planning horizon. The behavior of the \cmapf solver is illustrated in Figure~\ref{fig:snapshotBadExample}(b), where it will have $a_1$ move counter-clockwise towards its target, while $a_2$ attempts to reach its target by moving counter-clockwise. After two time steps, agent $a_1$ reaches its target and receives its next target, $s_1$. This causes $a_2$ to start moving clockwise, to avoid disrupting agent $a_1$. This repeats indefinitely, resulting in starvation of agent $a_2$, and a throughput of $\mathit{timesteps}\div2$. This will be true for any horizon larger than $6$.
Conversely, the \tmapf solver will have $a_2$ wait in place for $a_1$ to reach $t_1$, as the solver is aware that $a_1$ will not be blocking $t_1$ indefinitely after reaching it. This is illustrated in Figure~\ref{fig:snapshotBadExample}(c.1). After agent $a_1$ reaches $t_1$ and receives its new target of $s_1$, both agents will start moving clockwise in unison, and reach their new targets after two time steps. They will then move counter-clockwise, and repeat this cycle indefinitely as seen in Figure~\ref{fig:snapshotBadExample}(c.2), resulting in a throughput of $\mathit{timesteps} - 1$ . The same will hold true for any horizon larger than $2$.
We also studied this behavior experimentally using \cmapf and \tmapf versions of CBS and \prp , as well as PIBT. We ran the LMAPF problem for 500 time steps with an infinite planning horizon, and found that the total runtime was 29,198ms for CBS, 33ms for CBSt, 137ms for \prp , 68ms for \prpt , and 80ms for PIBT. Throughput was 250 for CBS, 499 for CBSt, 250 for \prp , 499 for \prpt , and 499 for PIBT. This experiment demonstrates how \tmapf algorithms are useful even while employing existing mitigation for the MAPF-LAMPF mismatch, achieving a higher throughput while requiring less runtime.

\begin{table}[t!] 
\centering
\scriptsize
\setlength{\tabcolsep}{2pt}
\begin{tabular}{@{} rr | r r r r r | r r r r r @{}} \toprule & & \multicolumn{5}{c|}{empty-48-48} & \multicolumn{5}{c}{warehouse-20-40-10-2-1} \\ A &  T & LNS & LNSt & PrP & PrPt & PIBT & LNS & LNSt & PrP & PrPt & PIBT \\ \midrule 100 & 10 & 278 & \textbf{347} & 427 & \textbf{868} & 765 & 115 & \textbf{121} & 138 & \textbf{141} & 135 \\ 100 & 20 & 434 & \textbf{748} & 632 & \textbf{821} & 795 & 123 & \textbf{126} & 142 & \textbf{143} & 137 \\ 100 & 30 & 630 & \textbf{802} & 744 & \textbf{852} & 822 & 129 & \textbf{133} & 142 & \textbf{143} & 136 \\ 100 & 40 & 701 & \textbf{789} & 783 & \textbf{853} & 836 & 129 & 129 & 138 & 138 & 132 \\ \midrule 300 & 10 & 203 & \textbf{361} & 205 & \textbf{400} & 1398 & 287 & \textbf{314} & 306 & \textbf{350} & 362 \\ 300 & 20 & 505 & \textbf{612} & 539 & \textbf{1595} & 2046 & 363 & \textbf{382} & 386 & \textbf{408} & 370 \\ 300 & 30 & \textbf{921} & 694 & 1185 & \textbf{2444} & 2163 & 394 & \textbf{402} & 412 & \textbf{421} & 373 \\ 300 & 40 & 1106 & \textbf{1011} & 1560 & \textbf{2457} & 2239 & 384 & \textbf{392} & 402 & \textbf{407} & 367 \\ \midrule 500 & 10 & 101 & \textbf{217} & 91 & \textbf{225} & 1602 & 338 & \textbf{425} & 336 & \textbf{432} & 478 \\ 500 & 20 & 301 & \textbf{396} & 317 & \textbf{409} & 2557 & 492 & \textbf{543} & 491 & \textbf{548} & 515 \\ 500 & 30 & 414 & \textbf{434} & 445 & \textbf{495} & 2997 & 571 & \textbf{602} & 585 & \textbf{613} & 512 \\ 500 & 40 & 419 & \textbf{441} & 504 & \textbf{566} & 3192 & 615 & \textbf{623} & 625 & \textbf{637} & 513 \\ \bottomrule \end{tabular}

\caption{Average throughput in our LMAPF dense targets experiment. Columns A and T show the numbers of agents and targets, respectively.}
\label{tbl:denseLL}
\end{table}

\section{Conclusion and Future Work}

In this work, we proposed solving LMAPF problems by solving a sequence of MAPF variants that we call Transient MAPF (\tmapf), in which agents do not have to stay at their targets after visiting them. 
We described that such a variant is needed to address a mismatch between the requirements of LMAPF and MAPF, and showed how to adapt existing \cmapf algorithms to solve it. 
We evaluated using our \tmapf algorithms within RHCR to solve lifelong MAPF. 
Our results showed no significant advantage for using \tmapf algorithms when the targets are spread uniformly. However, when the targets are limited to a fixed small number of locations, using the \tmapf algorithms is important, often achieving higher system throughput than their \cmapf counterparts. 
Future work in this line of research may focus on finding ways to select where agents should end their paths in \tmapf to avoid disrupting the other agents. Another direction is to generalize \tmapf to applications where agents must spend some amount of time at their target to complete some task.

\bibliography{submission}

\newpage

\section{Supplementary Materials}

\begin{figure*}[t]
\centering
\includegraphics[width=0.24\textwidth]{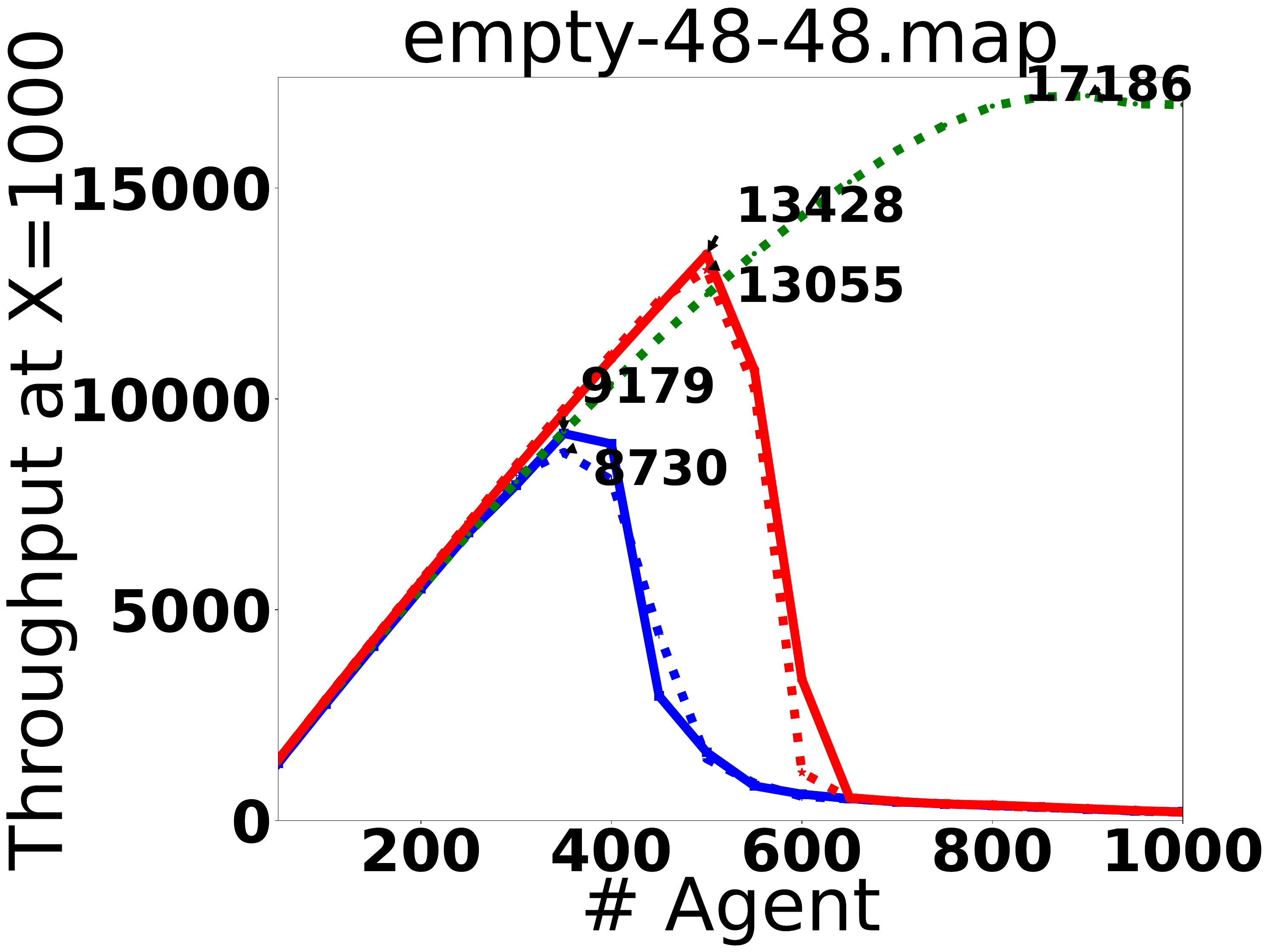}
\includegraphics[width=0.24\textwidth]{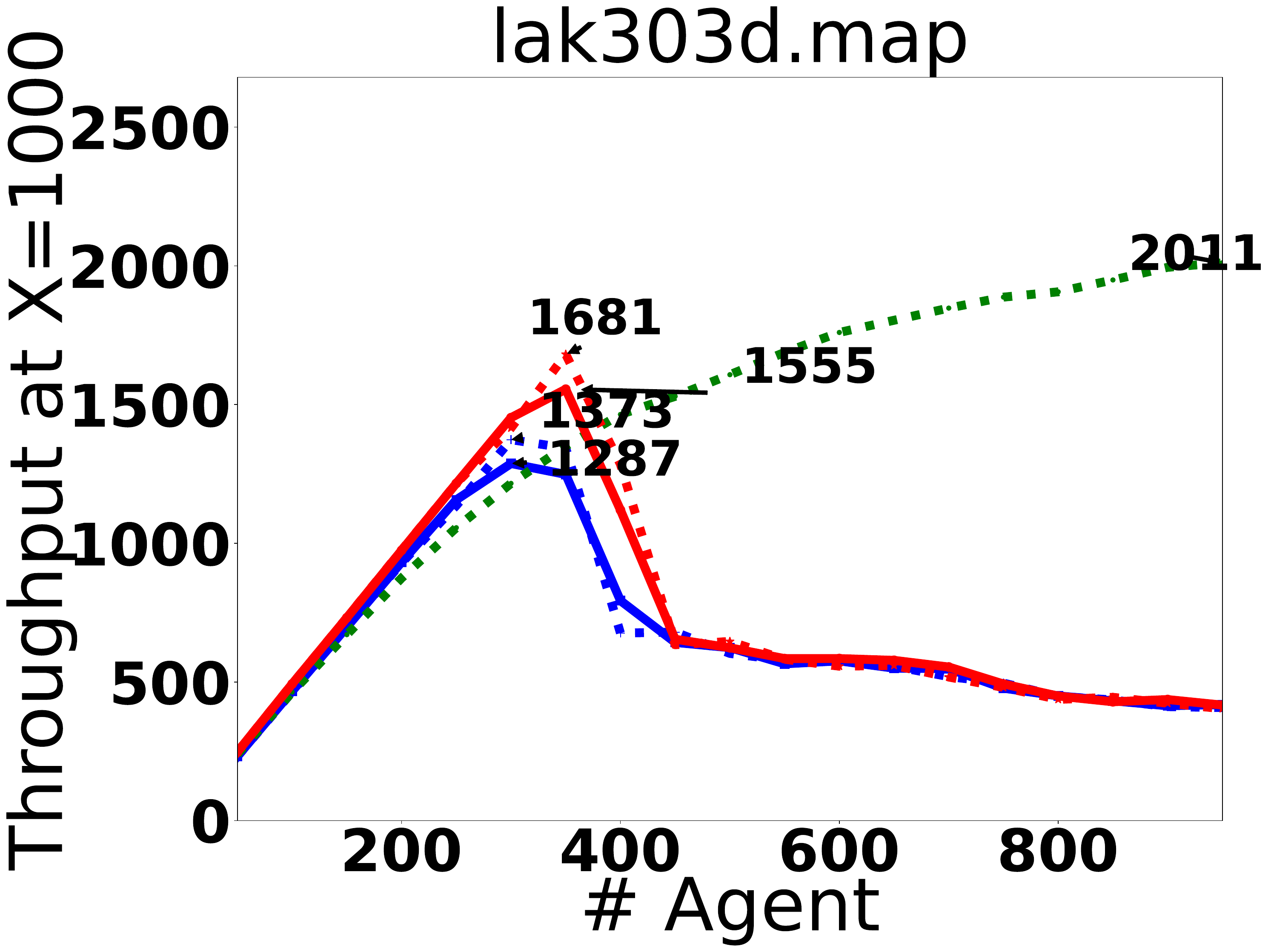}
\includegraphics[width=0.24\textwidth]{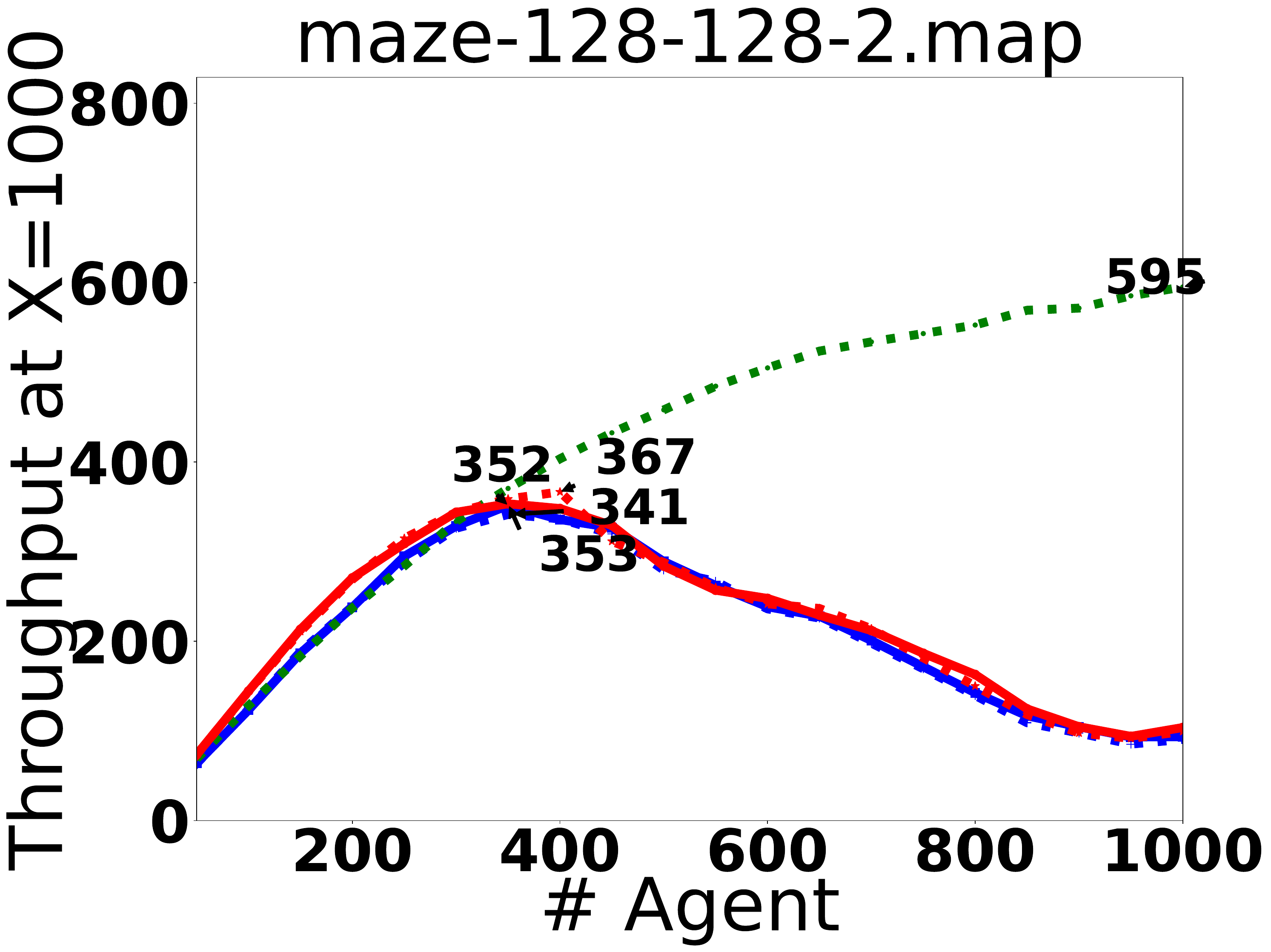}
\includegraphics[width=0.24\textwidth]{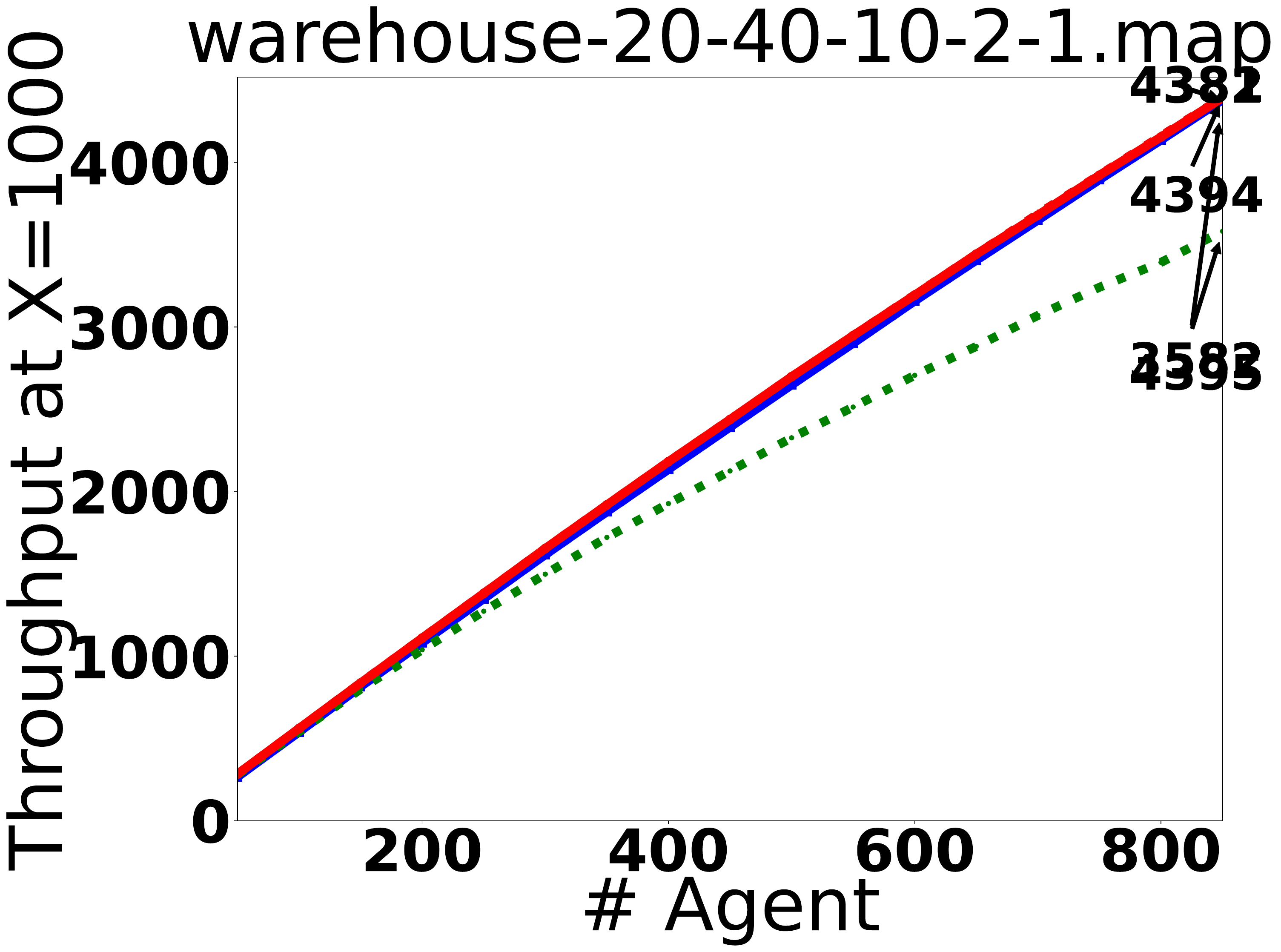}
\includegraphics[width=0.6\textwidth]{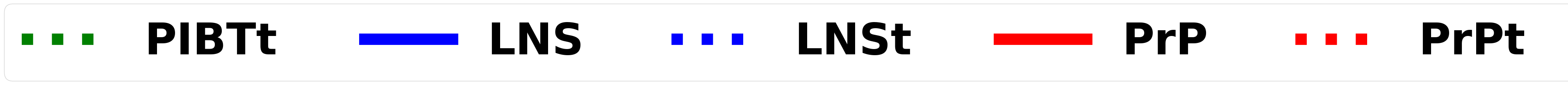}
\caption{Throughput achieved by the different algorithms as factor of the number of agents.} 
\label{fig:ThroughputAtT300}
\end{figure*}

Figure~\ref{fig:ThroughputAtT300} shows the throughput achieved by the different solvers on different maps within 300 time steps, averaged for each number of agents. The horizontal axis represents the number of agents in the problem instance and the vertical axis represents the throughput. 
We can see that in all maps the throughput achieved by \prp, \prpt, LNS, and \lnst was similar. PIBT, \lacam, and \lacamt also achieved thoughputs that were similar to each other. Which of the two groups had higher throughput depended on the specific map and the number of agents. 
For example, in \maze, the \prp group had higher throughput between 200 and 900 agents, whereas the \lacam group had higher throughput with 1000 agents.

\begin{table*}[t!] 
\centering
\small
\setlength{\tabcolsep}{4pt}
\begin{tabular}{cc|rrrrr|rrrrr}
\hline
          &            & \multicolumn{5}{c|}{empty-48-48}                 & \multicolumn{5}{c}{lak303d}                      \\ \hline
\# agents & \# targets & LNS  & LNSt  & PrP  & PrPt  & PIBTt & LNS   & LNSt & PrP  & PrPt & PIBTt \\ \hline
100       & 10         & 279  & 347   & 428  & 868   & 765   & 141   & 149  & 158  & 166  & 153   \\
100       & 20         & 434  & 748   & 633  & 822   & 796   & 124   & 126  & 134  & 135  & 125   \\
100       & 30         & 631  & 802   & 745  & 852   & 823   & 119   & 121  & 129  & 129  & 119   \\
100       & 40         & 701  & 790   & 783  & 854   & 836   & 119   & 119  & 127  & 127  & 118   \\ \hline
200       & 10         & 265  & 364   & 278  & 491   & 1238  & 249   & 282  & 279  & 328  & 273   \\
200       & 20         & 673  & 785   & 801  & 1610  & 1496  & 242   & 244  & 263  & 272  & 237   \\
200       & 30         & 1010 & 1207  & 1198 & 1678  & 1551  & 235   & 241  & 255  & 258  & 227   \\
200       & 40         & 1102 & 1465  & 1334 & 1678  & 1589  & 237   & 240  & 254  & 256  & 226   \\ \hline
300       & 10         & 204  & 361   & 206  & 400   & 1398  & 337   & 365  & 368  & 445  & 376   \\
300       & 20         & 506  & 612   & 540  & 1595  & 2047  & 344   & 355  & 380  & 402  & 331   \\
300       & 30         & 921  & 694   & 1185 & 2445  & 2164  & 343   & 351  & 377  & 385  & 318   \\
300       & 40         & 1106 & 1011  & 1560 & 2458  & 2240  & 346   & 348  & 375  & 377  & 315   \\ \hline
400       & 10         & 138  & 291   & 135  & 314   & 1513  & 360   & 361  & 365  & 392  & 459   \\
400       & 20         & 415  & 481   & 412  & 543   & 2388  & 413   & 407  & 463  & 451  & 411   \\
400       & 30         & 544  & 553   & 670  & 2371  & 2665  & 431   & 433  & 480  & 462  & 400   \\
400       & 40         & 654  & 572   & 885  & 3161  & 2799  & 432   & 442  & 480  & 467  & 395   \\ \hline
500       & 10         & 102  & 217   & 92   & 225   & 1603  & 344   & 326  & 344  & 332  & 526   \\
500       & 20         & 302  & 396   & 317  & 409   & 2557  & 437   & 410  & 441  & 434  & 485   \\
500       & 30         & 415  & 434   & 446  & 495   & 2997  & 467   & 452  & 474  & 456  & 471   \\
500       & 40         & 420  & 441   & 504  & 567   & 3192  & 481   & 462  & 488  & 473  & 465   \\ \hline
          &            & \multicolumn{5}{c|}{maze-128-128-2}              & \multicolumn{5}{c}{warehouse-20-40-10-2-1}       \\ \hline
\# agents & \# targets & LNS  & LNSt  & PrP  & PrPt  & PIBTt & LNS   & LNSt & PrP  & PrPt & PIBTt \\ \hline
100       & 10         & 23  & 23   & 24  & 24   & 22    & 116 & 122  & 139 & 142  & 136   \\
100       & 20         & 23  & 23   & 24  & 24   & 22    & 124 & 127  & 142 & 144  & 137   \\
100       & 30         & 23  & 23   & 25  & 25   & 23    & 129 & 133  & 143 & 143  & 136   \\
100       & 40         & 22  & 22   & 24  & 24   & 22    & 129 & 130  & 138 & 139  & 133   \\ \hline
200       & 10         & 46  & 47   & 48  & 48   & 42    & 225 & 240  & 253 & 273  & 251   \\
200       & 20         & 48  & 48   & 50  & 50   & 45    & 251 & 258  & 274 & 280  & 256   \\
200       & 30         & 48  & 48   & 50  & 50   & 45    & 261 & 265  & 281 & 284  & 259   \\
200       & 40         & 45  & 45   & 47  & 47   & 42    & 257 & 261  & 273 & 275  & 255   \\ \hline
300       & 10         & 70  & 69   & 71  & 72   & 62    & 287 & 315  & 307 & 351  & 362   \\
300       & 20         & 71  & 71   & 74  & 75   & 63    & 364 & 382  & 387 & 408  & 370   \\
300       & 30         & 73  & 73   & 76  & 76   & 65    & 394 & 402  & 413 & 421  & 373   \\
300       & 40         & 69  & 68   & 71  & 72   & 61    & 385 & 392  & 402 & 407  & 367   \\ \hline
400       & 10         & 92  & 93   & 94  & 95   & 80    & 313 & 376  & 322 & 388  & 465   \\
400       & 20         & 94  & 95   & 97  & 97   & 82    & 438 & 476  & 454 & 491  & 476   \\
400       & 30         & 95  & 96   & 99  & 99   & 83    & 501 & 517  & 515 & 538  & 478   \\
400       & 40         & 91  & 90   & 94  & 94   & 80    & 506 & 515  & 524 & 532  & 471   \\ \hline
500       & 10         & 112 & 113  & 115 & 115  & 94    & 338 & 426  & 336 & 433  & 566   \\
500       & 20         & 115 & 116  & 120 & 120  & 100   & 492 & 543  & 491 & 549  & 579   \\
500       & 30         & 118 & 118  & 120 & 120  & 101   & 572 & 603  & 586 & 614  & 576   \\
500       & 40         & 109 & 110  & 111 & 112  & 95    & 615 & 632  & 626 & 637  & 572   \\ \hline
\end{tabular}
\caption{Throughput achieved by different algorithms in our LMAPF dense targets experiment, averaged over 25 instances.}
\label{tbl:fullDenseLL}
\end{table*}

Table~\ref{tbl:fullDenseLL} shows the full results of the dense target experiments on maps \emptymap, \warehouse, \maze , and lak303d. 
For every map and number of agents, we highlighted in bold the algorithm that achieved the highest throughput between every \tmapf algorithm and its \cmapf counterpart.  
The results show that the \tmapf algorithms have a significant advantage in terms of throughput when the number of possible targets is limited. For example, the highest throughput achieved on \warehouse with 500 agents and 30 targets was 594, achieved by \lnst and \prpt. For the same map, a number of agents, and a number of targets, LNS and \prp yielded a throughput of only 361 and 352, respectively

\end{document}